\documentclass[aps,pra,twocolumn,floatfix,showpacs]{revtex4}
\usepackage[dvips]{color}
\usepackage{amsmath,amssymb,amsfonts,bm,graphicx,latexsym}
\usepackage[hypertex,breaklinks=true,colorlinks=false]{hyperref}

\newcommand{\ua}{\uparrow}
\newcommand{\da}{\downarrow}

\newcommand{\rv}{\bm{r}}

\newcommand{\pv}{\bm{p}}

\newcommand{\Lv}{\bm{L}}

\newcommand{\Av}{\bm{A}}

\newcommand{\Psih}{\hat{\Psi}}

\newcommand{\tot}{\mathrm{tot}}

\newcommand{\ev}{\bm{e}}
\newcommand{\Lambdav}{\bm{\Lambda}}

\newcommand{\PRL}[3]{Phys. Rev. Lett. {\bf #1}, \href{http://link.aps.org/abstract/PRL/v#1/e#2}{#2} (#3)}
\newcommand{\PRLp}[3]{Phys. Rev. Lett. {\bf #1}, \href{http://link.aps.org/abstract/PRL/v#1/p#2}{#2} (#3)}
\newcommand{\PRA}[3]{Phys. Rev. A {\bf #1}, \href{http://link.aps.org/abstract/PRA/v#1/e#2}{#2} (#3)}
\newcommand{\PRAp}[3]{Phys. Rev. A {\bf #1}, \href{http://link.aps.org/abstract/PRA/v#1/p#2}{#2} (#3)}
\newcommand{\PRAR}[3]{Phys. Rev. A {\bf #1}, \href{http://link.aps.org/abstract/PRA/v#1/e#2}{#2} (R) (#3)}
\newcommand{\PRB}[3]{Phys. Rev. B {\bf #1}, \href{http://link.aps.org/abstract/PRB/v#1/e#2}{#2} (#3)}
\newcommand{\PRBp}[3]{Phys. Rev. B {\bf #1}, \href{http://link.aps.org/abstract/PRB/v#1/p#2}{#2} (#3)}

\newcommand{\PRE}[3]{Phys. Rev. E {\bf #1}, \href{http://link.aps.org/abstract/PRE/v#1/e#2}{#2} (#3)}
\newcommand{\RMP}[3]{Rev. Mod. Phys. {\bf #1}, \href{http://link.aps.org/abstract/RMP/v#1/e#2}{#2} (#3)}
\newcommand{\arXiv}[1]{arXiv:\href{http://arxiv.org/abs/#1}{#1}}

\begin{document}

\title{
Global phase diagram of two-component Bose gases in antiparallel magnetic fields
}
\author{Shunsuke Furukawa}
\affiliation{Department of Physics, University of Tokyo, 7-3-1 Hongo, Bunkyo-ku, Tokyo 113-0033, Japan}
\author{Masahito Ueda}
\affiliation{Department of Physics, University of Tokyo, 7-3-1 Hongo, Bunkyo-ku, Tokyo 113-0033, Japan}
\date{\today}


\begin{abstract}
We study the ground-state phase diagram of two-dimensional two-component (or pseudospin-$\frac12$) Bose gases in mutually antiparallel synthetic magnetic fields  
in the space of the total filling factor and the ratio of the intercomponent coupling $g_{\ua\da}$ to the intracomponent one $g>0$. 
This time-reversal-invariant setting represents a bosonic analogue of spin Hall systems. 
Using exact diagonalization, we find 
that (fractional) quantum spin Hall states composed of a pair of nearly independent quantum Hall states 
are remarkably robust and persist for $g_{\ua\da}$ up to as large as $g$.
For $g_{\ua\da}=-g$, we find the exact many-body ground state in which particles in different spin states form pairs. 
This gives the exact critical line beyond which the system collapses. 
\end{abstract}

\maketitle


\section{Introduction}


Recent years have witnessed a rapid development in experimental techniques for creating synthetic gauge fields in ultracold atomic gases \cite{Dalibard11,Goldman13}. 
By optically coupling internal states of atoms, a nearly uniform synthetic magnetic field has been created \cite{Lin09}, 
opening up a new avenue towards the realization of quantum Hall (QH) states. 
Moreover, magnetic fields of mutually antiparallel directions have been optically generated
in two-component (pseudospin-$\frac12$) Bose gases, 
allowing observation of a spin Hall effect arising from spin-dependent Lorentz forces \cite{Beeler13}. 
There have also been proposals to realize similar gauge fields by inducing laser-assisted tunneling in tilted optical lattices \cite{Aidelsburger13,Kennedy13}. 
Although the spin Hall effect observed in Ref.~\cite{Beeler13} was still in a classical regime, the physical settings of Refs.~\cite{Beeler13,Aidelsburger13,Kennedy13}
show striking resemblances with quantum spin Hall (QSH) systems studied in semiconductors. 


The QSH effect was first studied in graphene \cite{Kane05} and semiconductors with a strain gradient structure \cite{Bernevig06}. 
It was later experimentally observed in HgTe/CdTe quantum wells \cite{Konig07} following a theoretical proposal \cite{Bernevig06_Science}. 
A notable feature of QSH systems is that they exhibit a pair of gapless edge modes protected by time-reversal symmetry 
while having an excitation gap in the bulk. 
The simplest model of QSH systems is a pair of integer QH systems with opposite chiralities. 
A natural generalization is to couple a pair of fractional QH states 
to construct an interacting analogue of QSH states with a fractionally quantized spin Hall conductance \cite{Bernevig06}. 
Interacting two-component atomic gases in high antiparallel synthetic magnetic fields would provide an ideal platform 
for studying correlated quantum phenomena in spin Hall systems. 
While fractional QSH states would appear naturally for a strong intracomponent repulsion $g>0$ and a weak intercomponent coupling $g_{\ua\da}$ \cite{Liu09}, 
it is interesting to ask whether they survive or are replaced by new quantum phases when $g_{\ua\da}/g$ is increased. 

\begin{figure}
\begin{center}
\includegraphics[width=0.43\textwidth]{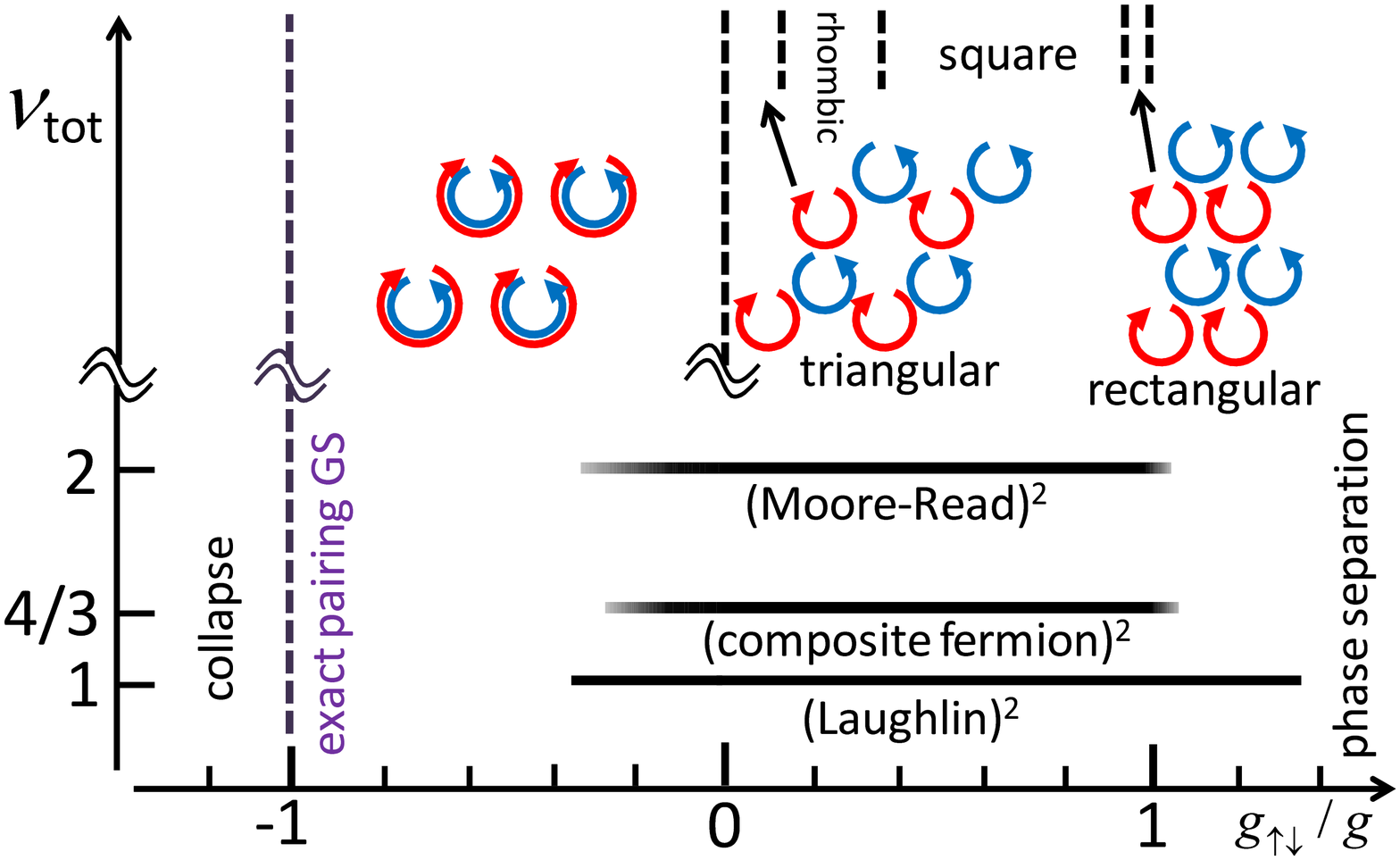}
\end{center}
\caption{(color online) 
Ground-state phase diagram in the space of the total filling factor $\nu_\tot=(N_\ua+N_\da)/N_\phi$ and the ratio of the intercomponent to intracomponent coupling constants,  $g_{\ua\da}/g$. 
Here $N_\ua~(N_\da)$ is the number of $\ua(\da)$-spin particles, 
and $N_\phi$ is the number of magnetic flux quanta piercing each component.
Within the Gross-Pitaevskii mean-field theory, a large-$\nu_\tot$ region exhibits the same vortex phase diagram as in the case of parallel magnetic fields 
studied previously \cite{Mueller02,Kasamatsu03}. 
For small $\nu_\tot$, however, (fractional) quantum spin Hall (QSH) states 
composed of a pair of nearly independent quantum Hall states (Laughlin, composite fermion, and Moore-Read states) 
appear over wide ranges of $g_{\ua\da}/g$ (see horizontal bars) 
in dramatic contrast to the case of parallel magnetic fields where SU(2)-symmetric quantum Hall states emerge 
for $g_{\ua\da}\approx g$ at $\nu_\tot=4/3$ \cite{Hormozi12,Grass12,FurukawaUeda12,Wu13} and $2$ \cite{Senthil13,FurukawaUeda13,Wu13,Regnault13}. 
For $\nu_\tot=1$, the range of the QSH phase (solid bar) is determined through an appropriate extrapolation to the thermodynamic limit. 
For $\nu_\tot=4/3$ and $2$, such an extrapolation could not be taken, 
and our best estimates (shaded bars) are based on the results for the largest system sizes in our calculations. 
On the critical line $g_{\ua\da}/g=-1$ beyond which the system collapses \cite{Comment_collapse}, 
pairing states constitute the exact many-body ground states. 
}
\label{fig:phase}
\end{figure}


In this paper, we study two-dimensional (2D) pseudospin-$\frac12$ Bose gases in high antiparallel synthetic magnetic fields. 
We determine the ground-state (GS) phase diagram 
in the space of the total filling factor $\nu_\tot=(N_\ua+N_\da)/N_\phi$ 
and the ratio $g_{\ua\da}/g$ (see Fig.~\ref{fig:phase}). 
Here, $N_\phi$ is the number of magnetic flux quanta piercing each component, $N_\ua$ ($N_\da$) is the number of particles in the spin state $\ua$ ($\da$).  
We will assume balanced populations $N_\ua=N_\da$ in most of the calculations. 


For $g_{\ua\da}=0$, the system decouples into two independent scalar Bose gases. 
The single-component problem in a synthetic magnetic field can be implemented in a rotating gas 
and has been studied in a number of works \cite{Cooper08_review}. 
For moderate magnetic fields, Abrikosov's triangular vortex lattice is formed in a Bose-Einstein condensate as observed experimentally \cite{AboShaeer01}. 
For high magnetic fields (i.e., small filling factors $\nu=N/N_\phi$), theory predicts that the vortex lattice melts, 
and incompressible QH states appear at various integer and fractional $\nu(\lesssim 6)$. 
Examples include a bosonic Laughlin state at $\nu=1/2$ \cite{Wilkin98}, a composite fermion state at $\nu=2/3$ \cite{Regnault04}, 
and a Moore-Read Pfaffian state \cite{Moore91} at $\nu=1$ \cite{Cooper01}. 
The coupling between two Bose gases in {\it parallel} magnetic fields leads to an even richer variety of phases as studied 
in both regimes of vortex lattices \cite{Mueller02,Kasamatsu03} and QH states \cite{Hormozi12,Grass12,FurukawaUeda12,Senthil13,FurukawaUeda13,Wu13,Regnault13}.
We will make comparisons between the cases of parallel and antiparallel magnetic fields in the course of our analyses. 


Our main results for pseudospin-$\frac12$ Bose gases in {\it antiparallel} magnetic fields are summarized in Fig.~\ref{fig:phase}. 
For large $\nu_\tot~(\gg 1)$, we show that within the Gross-Pitaevskii (GP) mean-field theory, the system shows exactly the same vortex structures 
as in the case of parallel magnetic fields studied previously \cite{Mueller02,Kasamatsu03}. 
For small $\nu_\tot$, we find through exact diagonalization calculations 
that (fractional) QSH states that are well approximated by a pair of independent QH states 
are markedly stable and survive for $g_{\ua\da}$ up to as large as $g$. 
It is remarkable that the two components remain nearly uncorrelated for such large intercomponent couplings. 
While similar bosonic systems have also been studied previously \cite{Liu09,Fialko14}, 
our calculations provide a more systematic determination of the parameter ranges of the QSH states \cite{Comment_Fialko14}. 
For $g_{\ua\da}=-g$, we find the exact many-body GS with a novel paring structure. 


Here we comment on related studies. 
A similar problem of two coupled fractional QH states with opposite chiralities 
have been studied in models of interacting spin-$\frac12$ fermions in lattices \cite{Neupert11} and continuous space \cite{ChenYang12}, 
and in a model of strained graphene \cite{Ghaemi12}. 
Compared with lattice systems \cite{Neupert11,Ghaemi12}, 
our simple setting can provide a clearer picture of the interplay between intracomponent and intercomponent interactions. 
Reference~\cite{ChenYang12} focuses on $\nu_\tot=2/3$, where two fermionic Laughlin states are coupled; 
our phase diagram for $\nu_\tot=1$ (corresponding to two coupled bosonic Laughlin states) has some similarities with the result of Ref.~\cite{ChenYang12}. 
Strong intercomponent interactions in spin Hall systems 
may potentially lead to novel time-reversal invariant topological states
(beyond simple product states) as proposed in Refs.~\cite{Bernevig06,Levin12,Santos11,Chen12,Lu12}.  
However, none of such phases have been found in our extensive numerical search in the present model 
(see Sec.~\ref{sec:incompress}). 
The stability of gapless edge states in coupled QH systems in the presence of intercomponent tunneling 
has also been discussed \cite{Levin09,Neupert11,Cappelli13,KochJanusz14}.  
Although this is an interesting problem, 
we focus on bulk properties in the absence of such tunneling  
and do not address edge properties in this paper. 


The rest of the paper is organized as follows. 
In Sec.~\ref{sec:GP},  we describe the setting of our system, and discuss the mean-field phase diagram for large $\nu_\tot(\gg 1)$. 
In Sec.~\ref{sec:ED}, we present our exact diagonalization analysis for small $\nu_\tot$. 
In particular, we perform an extensive search for incompressible states in the present model, 
and determine the parameter ranges of the QSH states at $\nu_\tot=1$, $4/3$, and $2$ in Fig.~\ref{fig:phase}. 
In Sec.~\ref{sec:exact_gs}, we derive some exact results on the point $g_{\ua\da}=-g$. 
In Sec.~\ref{sec:conclude}, we present a summary and an outlook for future studies. 
In Appendix \ref{app:LLL_sphere}, we describe some technical details related to numerical calculations in a spherical geometry in Sec.~\ref{sec:exact_gs}. 

\section{Mean-field theory for large filling factors} \label{sec:GP}

We consider a system of a 2D pseudospin-$\frac12$ Bose gas (in the $xy$ plane) 
subject to antiparallel magnetic fields $+ B$ and $-B$ along the $z$ axis for spin states $\alpha=\uparrow$ and $\downarrow$, respectively. 
We introduce the fictitious charge $q$ of a particle, and assume $qB>0$. 
We denote the strengths of the intracomponent and intercomponent contact interactions by $g$ and $g_{\ua\da}$, respectively. 
In the second-quantized form, the interaction Hamiltonian is written as
\begin{equation}\label{eq:Hint}
 H_\mathrm{int} 
 = \sum_{\alpha,\beta=\ua,\da} \frac{g_{\alpha\beta}}{2} \int d^2\bm{r} \Psih_\alpha^\dagger(\rv) \Psih_\beta^\dagger(\rv) \Psih_\beta(\rv) \Psih_\alpha(\rv),
\end{equation}
where $\Psih_\alpha (\rv)$ is the bosonic field operator for the spin state $\alpha$. 
We set $g_{\ua\ua}=g_{\da\da}\equiv g>0$ and $g_{\ua\da}=g_{\da\ua}$.
For a 2D system of area $A$, the number of magnetic flux quanta piercing each component is given by 
$N_\phi=|qB|A/(2\pi\hbar)=A/(2\pi \ell^2)$, where $\ell=\sqrt{\hbar/|qB|}$ is the magnetic length. 
Strongly correlated physics is expected to emerge when $N_\phi$ becomes comparable with or larger than the total number of particles, $N=N_\ua+N_\da$. 
We investigate the GS phase diagram of the system in the space of the total filling factor $\nu_\tot=N/N_\phi$ and the ratio $g_{\ua\da}/g$.

We first focus on the regime of large $\nu_\tot (\gg 1)$, 
where the system is expected to be well described by the Gross-Pitaevskii (GP) mean-field theory. 
In the GP theory, the field operators $\Psih_\alpha(\rv)$ in Eq.~\eqref{eq:Hint} are replaced by 
the condensate wave functions $\Psi_\alpha (\rv)$ which are determined 
by minimizing the GP energy functional
\begin{equation}
\begin{split}
E[\Psi_\uparrow,\Psi_\downarrow]=\int d^2 \bm{r} &
\Big[ \Psi_\ua^* {\cal K}(B) \Psi_\ua + \Psi_\da^* {\cal K}(-B) \Psi_\da\\
&+ \sum_{\alpha,\beta} \frac{g_{\alpha\beta}}{2} |\Psi_\alpha|^2|\Psi_\beta|^2\Big],
\end{split}
\end{equation}
where ${\cal K}(\pm B)$ is the single-particle Hamiltonian in the Landau gauge: 
\begin{equation}
{\cal K} (\pm B)=\frac{1}{2M} \left[ (-i\hbar\partial_x\pm qBy)^2+(-i\hbar\partial_y)^2 \right]. 
\end{equation}
Noting 
\begin{equation}
\int d^2 \bm{r} \Psi_\downarrow^* {\cal K}(-B) \Psi_\downarrow 
= \int d^2 \bm{r} \Psi_\downarrow {\cal K}(B) \Psi_\downarrow^*, 
\end{equation}
we find that the present system reduces to the case of parallel magnetic fields by the replacement $\Psi_\da\to\Psi_\da^*$. 
Namely, the GP energy functionals for the cases of parallel and antiparallel magnetic fields are related as 
\begin{equation}
E_\mathrm{antiparallel}[\Psi_\ua,\Psi_\da]=E_\mathrm{parallel}[\Psi_\ua,\Psi_\da^*].
\end{equation} 
This implies that the GS phase diagram for one case 
can be obtained from that for the other through time reversal of the $\da$ component.
The mean-field phase diagram for parallel magnetic fields is known \cite{Mueller02,Kasamatsu03} and summarized as follows.  
When $g_{\ua\da}=0$, each component independently forms a triangular vortex lattice. 
The two triangular lattices overlap for $g_{\ua\da}<0$ and are displaced from each other for small $g_{\ua\da}/g>0$. 
When the ratio of the coupling constants is increased in the range $0<g_{\ua\da}/g<1$, 
the two interlocked lattices undergo phase transitions from triangular to rectangular configurations 
(by taking a complex conjugate for the $\da$ component in these states, we obtain the vortex lattices as schematically illustrated in Fig.~\ref{fig:phase}). 
For $g_{\ua\da}/g>1$, the system displays exotic metastable states, such as double-core vortex lattices, stripes, and vortex sheets. 
All these vortex structures are expected to emerge (with the reversed current density for the $\da$ component) in our time-reversal invariant system as well. 


\section{Exact diagonalization analysis for small filling factors}\label{sec:ED}

In this section, we consider the regime of $\nu_\tot=O(1)$, 
where the system is expected to be strongly correlated. 
We assume that $B$ is so large that the interaction energy is much smaller than the Landau-level spacing $\hbar |qB|/M$. 
In this case, the restriction of the Hilbert space to the lowest Landau level (LLL) is legitimate. 
Within this restricted subspace, we have performed an exact diagonalization analysis of the Hamiltonian \eqref{eq:Hint}. 
We demonstrate that QSH states composed of two nearly independent QH states are remarkably robust 
and persist for $g_{\ua\da}$ up to as large as $g$ at $\nu_\tot=1$, $4/3$, and $2$ (see horizontal lines in Fig.~\ref{fig:phase}). 
Our results also indicate that in the present system, increasing $g_{\ua\da}/g$ only gradually diminishes the energy gaps of such QSH states 
and does not lead to any novel incompressible state (beyond simple product states) proposed in Refs.~\cite{Bernevig06,Levin12,Santos11,Chen12,Lu12}. 
In our analysis, we set 
\begin{equation}
 (g,g_{\ua\da})=G\ell^2 (\cos\gamma,\sin\gamma)
\end{equation}
with $G>0$, and change $\gamma$ in the range $-\frac{\pi}{2}\le \gamma\le \frac{\pi}{2}$. 

\subsection{Spherical geometry} \label{sec:sphere}

For our numerical calculations, we have employed a spherical geometry \cite{Haldane83,Fano86}, 
which has no edge and is therefore suitable for studying bulk properties around the center of a trapped gas. 
Here we briefly describe the basic setup for this geometry. 
Further details on this geometry such as the representation of the Hamiltonian in the LLL basis 
and the numerical methods used for diagonalizing the Hamiltonian are explained in Appendix~\ref{app:LLL_sphere}. 

We introduce the polar coordinates $(r,\theta,\phi)$ and the associated unit vectors $\ev_r$, $\ev_\theta$, $\ev_\phi$. 
We place magnetic monopoles of charges $\pm N_\phi (2\pi \hbar/q)$ with integer $N_\phi\equiv 2S$ 
at the center of the sphere.   
These monopoles produce magnetic fields $+ B \bm{e}_r$ and $- B \bm{e}_r$ for the $\ua$ and $\da$ particles, respectively, 
which reside on the sphere of radius $R=\ell \sqrt{S}$ with $\ell=\sqrt{\hbar/|q B|}$. 
We assume $qB>0$ in the following. 
Because of the spherical symmetry, many-body eigenstates can be classified by the total angular momentum $L$. 
For incompressible states on finite spheres, 
the relation between $N$ and $N_\phi$ involves a characteristic shift $\delta$:
\begin{equation}\label{eq:N_Nphi}
 N = \nu_\tot (N_\phi+\delta), 
\end{equation}
where $\delta$ depends on individual candidate wave functions. 
Incompressible GSs appear in general in the sector with $L=0$ for $(N_\phi,N)$ satisfying the relation \eqref{eq:N_Nphi}. 
The QSH states at $\nu_\tot=1, 4/3, 2$ in Fig.~\ref{fig:phase} have $\delta=2,3,2$, respectively. 

\subsection{Numerical search for incompressible states} \label{sec:incompress}

\begin{figure*}
\begin{center}
\includegraphics[width=0.48\textwidth]{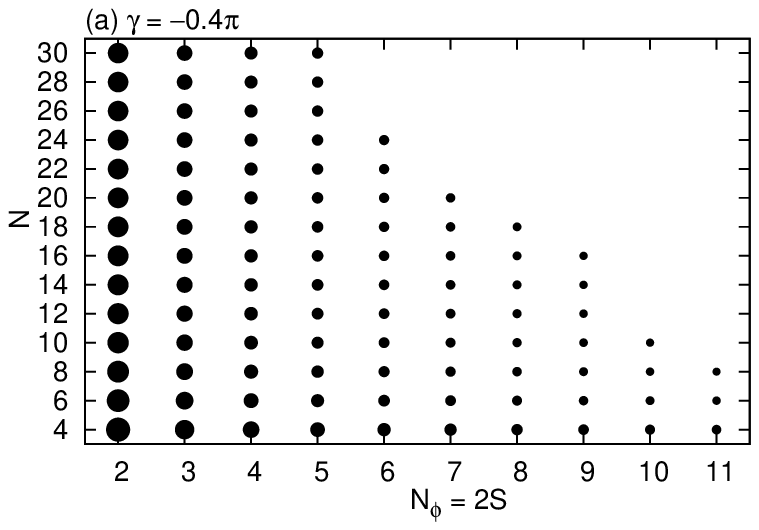}
\includegraphics[width=0.48\textwidth]{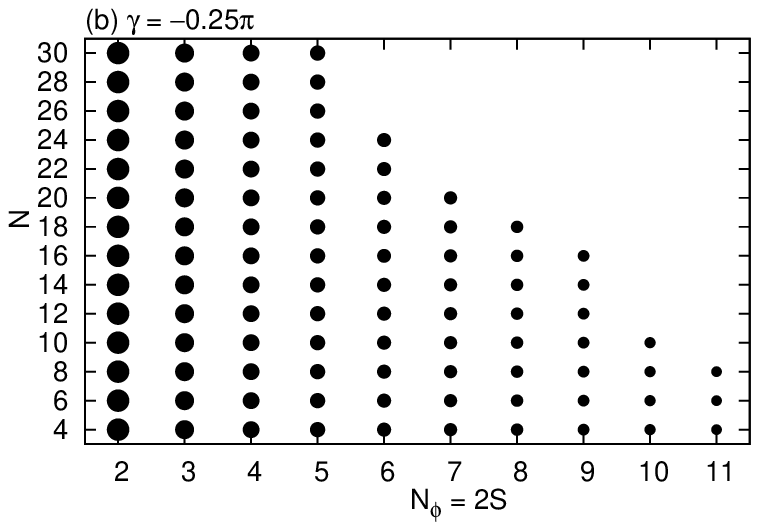}\\
\includegraphics[width=0.48\textwidth]{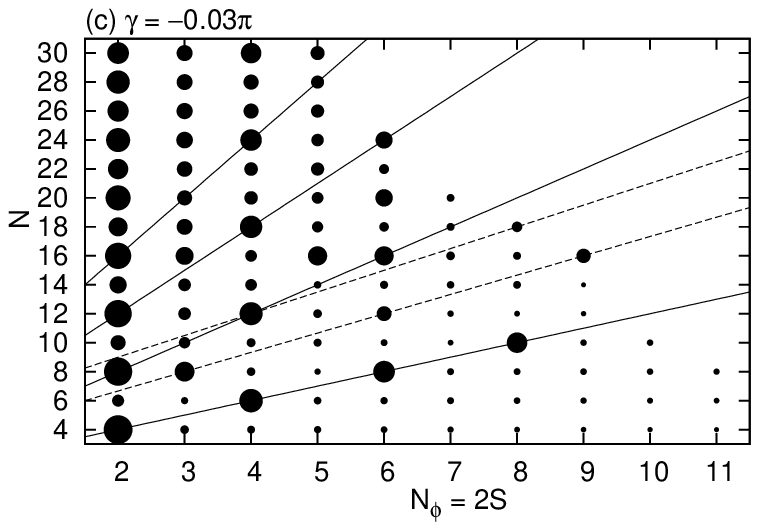}
\includegraphics[width=0.48\textwidth]{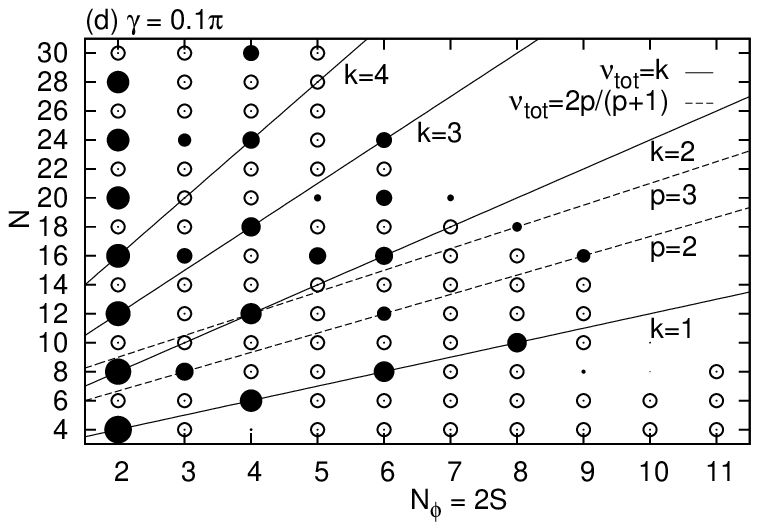}\\
\includegraphics[width=0.48\textwidth]{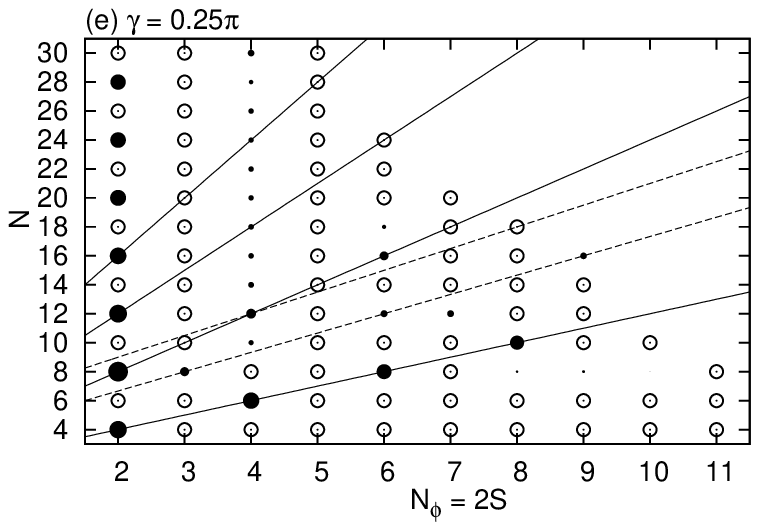}
\includegraphics[width=0.48\textwidth]{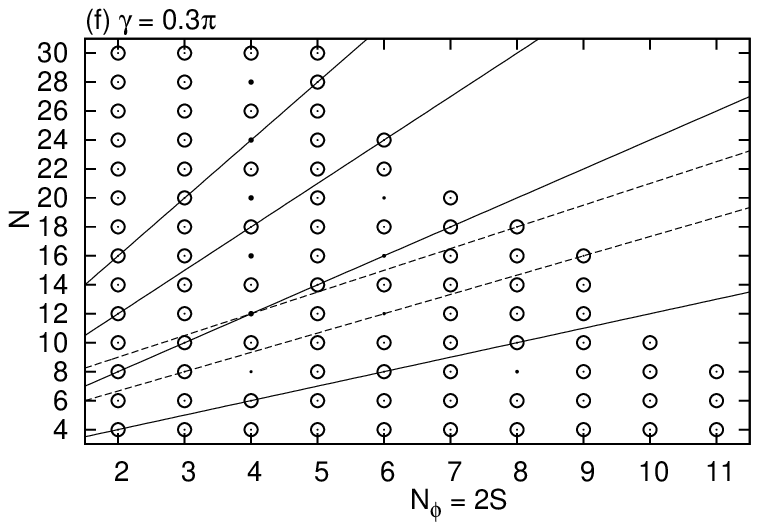}
\end{center}
\caption{
Candidates for incompressible GSs in the $(N_\phi,N)$ plane, 
calculated on a spherical geometry for different values of $\gamma=\arctan(g_{\ua\da}/g)$. 
Filled circles indicate GSs with the total angular momentum $L=0$, where incompressible states can appear; 
the area of each filled circle is proportional to the neutral gap $\Delta_n$. 
Empty circles indicate the GSs with $L>0$. 
Solid and broken lines indicate the relation \eqref{eq:N_Nphi} for $(\nu_\tot,\delta)=(k,2)$ and $(\frac{2p}{p+1},p+1)$, respectively. 
The QSH states composed of a pair of Laughlin, composite fermion, and Moore-Read states appear for $(\nu_\tot,\delta)=(1,2)$, $(\frac43,3)$, and $(2,2)$, respectively. 
Data points are missing for large $N_\phi$ or $N$ due to an exponentially increasing computation time. 
}
\label{fig:gsLtri}
\end{figure*}

Through exact diagonalization calculations on a spherical geometry, 
we have carried out an extensive search for incompressible GSs in the $(N_\phi,N)$ plane 
for different values of $\gamma=\arctan(g_{\ua\da}/g)$; see Fig.~\ref{fig:gsLtri}.  
Incompressible states in general appear as the unique GSs with $L=0$, 
which are indicated by filled circles in Fig.~\ref{fig:gsLtri}. 
The area of each filled circle is proportional to the neutral gap $\Delta_n$, 
which is defined as the excitation gap for fixed $(N_\phi,N_\ua,N_\da)$. 
Solid lines indicate the relation \eqref{eq:N_Nphi} for $(\nu_\tot,\delta)=(k,2)$ with $k=1,2,3,4$; 
they correspond to the QSH states made of two Read-Rezayi states \cite{Read99}
and include in particular the cases of two Laughlin ($k=1$) or Moore-Read $(k=2)$ states 
(we note that these states appear only for even $N_\phi$). 
Broken lines indicate the relation \eqref{eq:N_Nphi} for $(\nu_\tot,\delta)=\left( \frac{2p}{p+1},p+1 \right)$ with $p=2,3$;  
they correspond to the QSH states made of two composite fermion states in Jain's principal sequence. 

We first analyze the case of $0<\gamma<\frac{\pi}2$. 
For (d) $\gamma=0.1\pi$ and (e) $\gamma=0.25\pi$, we find that the $L=0$ GSs appear for $(N_\phi,N)$ on the solid and broken lines. 
However, the values of the neutral gap are much smaller in (e). 
For (f) $\gamma=0.3\pi$, most of the $L=0$ GSs disappear; even when they survive, the gap values are very small. 
Comparing (e) with (d), we do not find the emergence of any new $L=0$ GS with an appreciable energy gap above it. 
This indicates that within the present model, increasing $g_{\ua\da}/g$ only gradually diminishes the energy gaps of QSH states 
made of two QH states and does not lead to any new incompressible state. 
A possible emergence of a new incompressible state in the presence of other perturbations such as the introduction of an optical lattice or longer-range interactions
is an interesting future problem. 

We next analyze the case of $-\frac{\pi}2<\gamma<0$. 
In this case, $L=0$ GSs appear for all $(N_\phi,N)$ we have investigated. 
However, it is important to analyze whether the energy gaps above these GSs remain nonvanishing in the thermodynamic limit. 
For (c) $\gamma=-0.03\pi$, the energy gaps are relatively large on the solid and broken lines, 
and finite energy gaps possibly remain on these lines in the thermodynamic limit. 
The energy gaps for other $(N_\phi,N)$ tend to decrease as we increase $N_\phi$. 
Remarkably, for (b) $\gamma=-0.25\pi$, the energy gaps do not depend on $N$ at all, 
and decrease monotonically as a function of $N_\phi$. 
At this point, the exact expression for the excitation energy can be found as we explain later; see Eq.~\eqref{eq:gap_J_1} in Sec.~\ref{sec:exact_es}. 
Since the energy gap vanishes as $N_\phi\to \infty$, no incompressible state appears. 
The energy gaps for (a)~$\gamma=-0.4\pi$ also shows a tendency to vanish as $N_\phi\to \infty$. 
At this parameter point, the system is not stable and spontaneously contracts as explained in Sec.~\ref{sec:collapse}. 

This subsection has focused on a global picture of the types and the ranges of incompressible states present in the model. 
More precise determination of the parameter ranges of the QSH phases requires a more detailed analysis, 
which we present next. 

\subsection{Determination of the parameter ranges of QSH states} \label{sec:determine_phase}

Here we present our numerical results for $\nu_\tot=1$, $4/3$, and $2$. 
For $g_{\ua\da}=0$, the GSs at these filling factors are given by QSH states 
composed of two independent Laughlin, composite fermion, or Moore-Read states, 
which have $\delta=2$, $3$, and $2$, respectively. 
We are particularly interested in the stability of such QSH states in the presence of the intercomponent coupling $g_{\ua\da}$. 
We also find an indication of a phase transition at $g_{\ua\da}/g=-1$, which is analyzed in more detail in Sec.~\ref{sec:collapse}. 

\begin{figure}
\begin{center}
\includegraphics[width=0.48\textwidth]{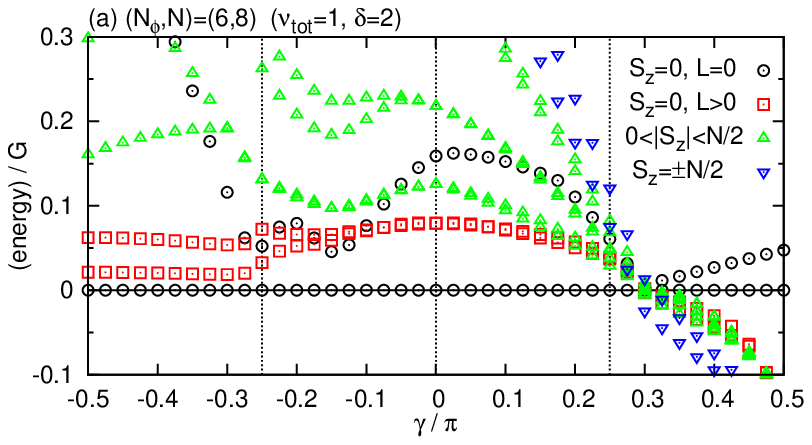}\\
\includegraphics[width=0.48\textwidth]{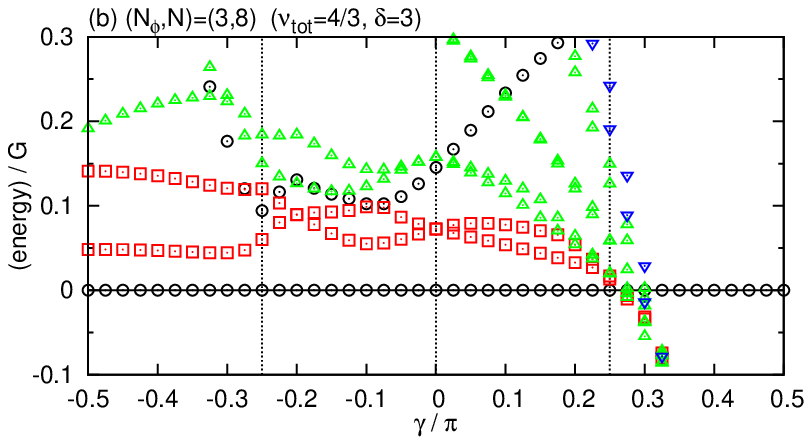}\\
\includegraphics[width=0.48\textwidth]{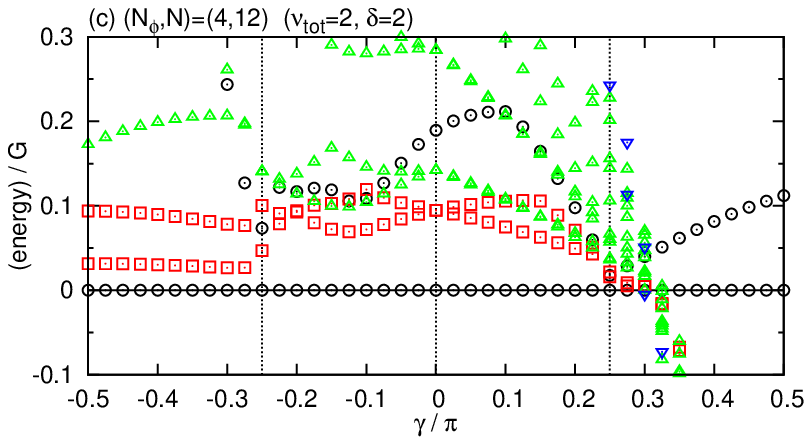}
\end{center}
\caption{(color online) 
Energy spectra versus $\gamma/\pi=\arctan(g_{\ua\da}/g)/\pi$ 
for (a)~$(N_\phi,N)=(6,8)$,  (b)~$(N_\phi,N)=(3,8)$, and (c)~$(N_\phi,N)=(4,12)$. 
The eigenstates are classified by $S_z=(N_\ua-N_\da)/2$ and the total angular momentum $L$. 
The two lowest eigenenergies in each sector of the Hilbert space are displayed. 
The lowest eigenenergy in the sector $(S_z,L)=(0,0)$ is subtracted from the entire spectrum. 
Vertical dotted lines indicate the cases of $g_{\ua\da}/g=0,\pm 1$. 
}
\label{fig:spectra}
\end{figure}

Figure~\ref{fig:spectra} displays energy spectra as a function of $\gamma/\pi=\arctan(g_{\ua\da}/g)/\pi$
for some $(N_\phi,N)$ corresponding to $(\nu_\tot,\delta)=(1,2)$, $(4/3,3)$, and $(2,2)$. 
While we are interested in the population-balanced case $N_\ua=N_\da$, 
we perform calculations for the imbalanced case $N_\ua\ne N_\da$ as well 
to examine a possible instability towards a phase separation. 
The eigenstates are classified by $S_z\equiv (N_\ua-N_\da)/2$ and $L$. 
In all the cases presented, a finite energy gap appears above the unique $L=0$ GS at $\gamma=0$ 
as expected for incompressible states; 
the gap value $\Delta/G\approx0.08$ for (a)~$(N_\phi,N)=(6,8)$ coincides with that calculated for a scalar gas with $4$ particles \cite{Regnault04}. 
The gap decreases monotonically as we increase $\gamma/\pi$. 
The GS stays in the sector $(S_z,L_\tot)=(0,0)$ for $\gamma/\pi<1/4$. 
Slightly above $\gamma/\pi=1/4$, the GS is replaced by the maximally imbalanced states with $S_z=\pm N/2$. 
This indicates that even for balanced populations $N_1=N_2$, 
the system locally favors formation of a single-component domain, leading to a phase separation. 
The level crossing point $\gamma_{c3}(N)$ gives an estimate of the transition point. 
Around $\gamma/\pi\approx -0.25$ and $-0.1$, some energy levels show extrema. 
In fact, phase transitions occur around these points, as we discuss next. 


\begin{figure}
\begin{center}
\includegraphics[width=0.48\textwidth]{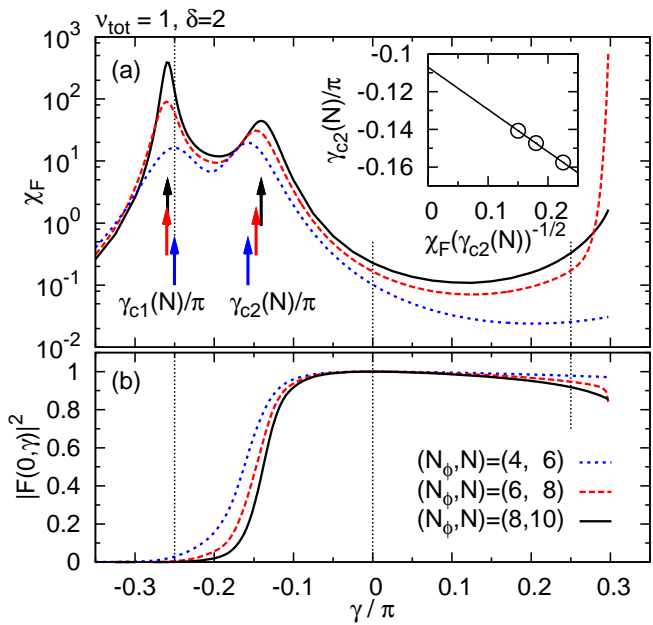}
\end{center}
\caption{(color online) 
(a) Fidelity susceptibility $\chi_F(\gamma)$ as a function of $\gamma/\pi$. 
The stability of a fractional QSH state composed of two Laughlin states is examined 
for three system sizes $(N_\phi,N)$ with filling factor $\nu_{\rm tot}=2$ and shift $\delta=2$. 
Peak positions indicated by arrows give finite-size estimates of the transition points. 
The inset shows an extrapolation of $\gamma_{c2}(N)/\pi$ to the thermodynamic limit 
using the scaling relation \eqref{eq:fid_scaling}. 
(b) Squared overlap with the decoupled case, i.e., $|F(0,\gamma)|^2$. 
Vertical dotted lines correspond to $\gamma=0,\pm \pi/4$. 
}
\label{fig:overlap}
\end{figure}


To detect phase transitions, 
we have calculated the fidelity susceptibility $\chi_F$ \cite{You07} as a function of $\gamma$. 
When two parameter points $\gamma$ and $\gamma+\delta\gamma$ are close enough, 
the overlap between the GSs at these points can be expanded as 
\begin{equation}
\begin{split}
 F(\gamma,\gamma+\delta\gamma)
 &\equiv |\langle\Psi(\gamma)|\Psi(\gamma+\delta\gamma)\rangle|\\
 &=1-\frac{\chi_F}{2}(\delta\gamma)^2+\dots, 
\end{split}
\end{equation}
which allows us to define the fidelity susceptibility 
\begin{equation} 
\chi_F(\gamma) = -2 \lim_{\delta\gamma\to 0} \frac{\ln F(\gamma,\gamma+\delta\gamma)}{(\delta\gamma)^2}.
\end{equation}
This quantity measures how rapidly the GS $|\Psi(\gamma)\rangle$ changes as a function of the model parameter $\gamma$. 

Let us first look at the result for $(\nu_\tot,\delta)=(1,2)$ in Fig.~\ref{fig:overlap}(a). 
For $\gamma<0$, we observe two peaks in $\chi_F(\gamma)$, which indicate phase transitions;
these peaks shift gradually and grow sharper with increasing $N$. 
The peak positions, $\gamma_{c1}(N)/\pi$ and $\gamma_{c2}(N)/\pi$ [arrows in Fig.~\ref{fig:overlap}(a)], 
can be used to make a finite-size estimation of the transition points. 
Using the scaling relation 
\begin{equation}\label{eq:fid_scaling}
\gamma_{c2}(N)-\gamma_{c2}(\infty)\simeq \mathrm{const.}\times \chi_F(\gamma_{c2}(N))^{-\frac12}
\end{equation}
with the peak height $\chi_F(\gamma_{c2}(N))$ \cite{Albuquerque10},  
we extrapolate the data of $\gamma_{c2}(N)$ to the thermodynamic limit, 
obtaining $\gamma_{c2}\approx -0.11\pi$ [inset of Fig.~\ref{fig:overlap}(a)]. 
The data of $\gamma_{c1}(N)$, by contrast, are located around $-\pi/4$ and do not depend smoothly on $N$; 
yet, we show later in Sec.~\ref{sec:collapse} and Sec.~\ref{sec:exact_pair_gs} that $\gamma_{c1}=-\pi/4$ gives the exact transition point in the thermodynamic limit.  
For $\gamma>0$, no peak appears in $\chi_F(\gamma)$ 
until the GS level crossing occurs at $\gamma_{c3}(N)\approx 0.30\pi$ (with little dependence on $N$) as in Fig.~\ref{fig:spectra}(a). 
These results indicate that a single phase is formed over 
\begin{equation}\label{eq:QSH_nu1}
-0.11\lesssim\gamma/\pi\lesssim 0.30 ~(-0.35\lesssim g_{\ua\da}/g\lesssim 1.4). 
\end{equation}
Figure~\ref{fig:overlap}(b) shows the squared overlap with the GS at $\gamma=0$: $|F(0,\gamma)|^2$. 
We find that $|F(0,\gamma)|^2$ indeed stays close to unity in the above range, 
indicating that a fractional QSH state, which is well approximated by the product of two independent Laughlin states, 
is realized over this range. 
The nature of the phase for $-0.25<\gamma/\pi\lesssim -0.11$ is currently unclear from the numerical data; 
yet the global phase structure in Fig.~\ref{fig:phase} suggests that overlapping triangular vortex lattices at large $\nu_\tot$ 
persist down to this small-$\nu_\tot$ regime. 

\begin{figure}
\begin{center}
\includegraphics[width=0.48\textwidth]{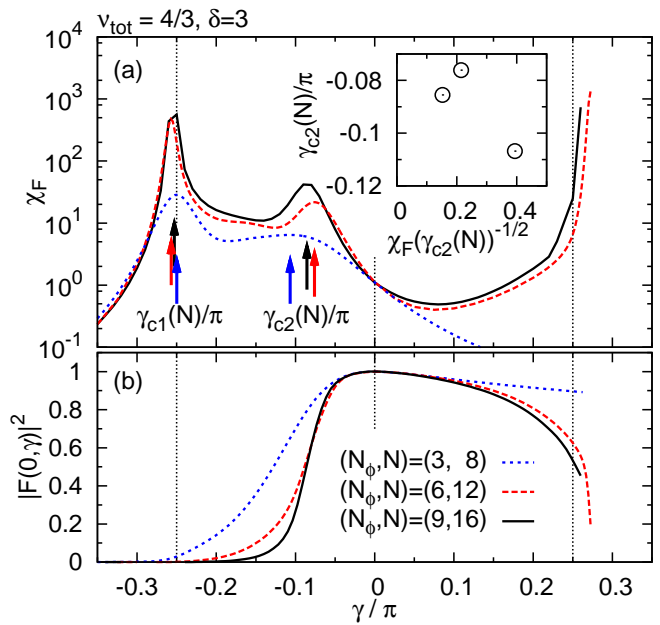}
\end{center}
\caption{(color online) 
(a) The fidelity susceptibility $\chi_F(\gamma)$ and (b) the squared overlap with the GS at $\gamma=0$, i.e., $|F(0,\gamma)|^2$, 
for $(\nu_\tot,\delta)=(4/3,3)$. 
The inset in (a) shows the relation between $\gamma_{c2}(N)/\pi$ and $\chi_F(\gamma_{c2}(N))^{-1/2}$, 
where the data points do not allow a smooth fit unlike the inset of Fig.~\ref{fig:overlap}(a). 
Arrows and vertical dotted lines are drawn in a manner similar to Fig.~\ref{fig:overlap}. 
}
\label{fig:overlap4o3}
\end{figure}

\begin{figure}
\begin{center}
\includegraphics[width=0.48\textwidth]{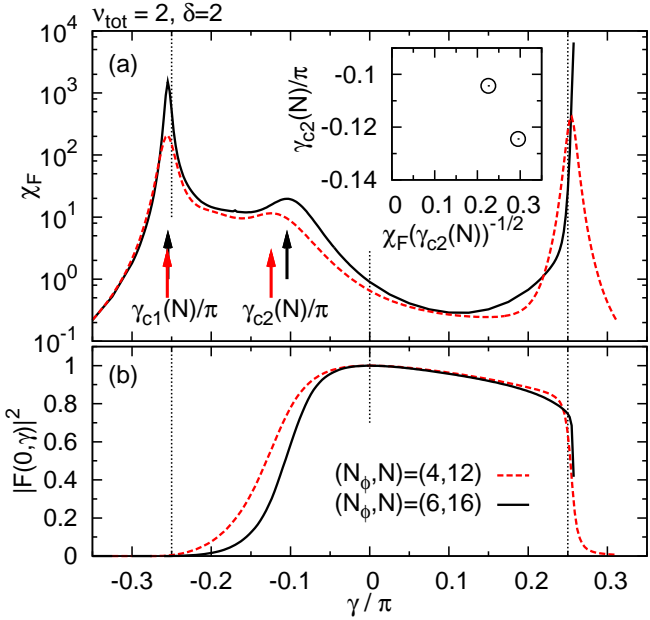}
\end{center}
\caption{(color online) 
(a) The fidelity susceptibility $\chi_F(\gamma)$ and (b) the squared overlap with the GS at $\gamma=0$, i.e., $|F(0,\gamma)|^2$, 
for $(\nu_\tot,\delta)=(2,2)$. 
The inset in (a) shows the relation between $\gamma_{c2}(N)/\pi$ and $\chi_F(\gamma_{c2}(N))^{-1/2}$; 
calculations for larger systems are required to make a reliable extrapolation to the thermodynamic limit as in the inset of Fig.~\ref{fig:overlap}(a). 
Arrows and vertical dotted lines are drawn in a manner similar to Fig.~\ref{fig:overlap}. 
}
\label{fig:overlap2}
\end{figure}

We have performed similar analyses for $(\nu_\tot,\delta)=(4/3,3)$ and $(2,2)$ as shown in Fig.~\ref{fig:overlap4o3} and \ref{fig:overlap2}. 
The fidelity susceptibility $\chi_F(\gamma)$ in Fig.~\ref{fig:overlap4o3}(a) and \ref{fig:overlap2}(a) shows two peaks indicating phase transitions. 
In contrast to the $\nu_\tot=1$ case, the data of $(\gamma_{c2}(N),\chi_F(\gamma_{c2}(N))^{-1/2})$ for the $\nu_\tot=4/3$ case [inset of Fig.~\ref{fig:overlap4o3}(a)] 
cannot be fitted by a smooth function. 
In the $\nu_\tot=2$ case, the calculations could be performed only for two system sizes. 
For both $\nu_\tot=4/3$ and $2$, the calculations for larger $N$ are not possible 
due to an exponentially increasing computation time. 
For these reasons, we cannot make an appropriate extrapolation of $\gamma_{c2}(N)$ to the thermodynamic limit in these cases.
We thus use $(\gamma_{c2}(N),\gamma_{c3}(N))$ for the largest $N$ for each $\nu_\tot$ 
as our best estimate of the range of the QSH phase, which is given by
\begin{equation}\label{eq:QSH_range}
\begin{split}
 \nu_\tot=\frac43: &-0.09\lesssim \gamma/\pi \lesssim 0.26 ~(-0.28\lesssim g_{\ua\da}/g \lesssim 1.06) \\
 \nu_\tot=2: &-0.10\lesssim \gamma/\pi \lesssim 0.26 ~(-0.34 \lesssim g_{\ua\da}/g \lesssim 1.05). 
\end{split}
\end{equation}
Figures \ref{fig:overlap4o3}(b) and \ref{fig:overlap2}(b) present the squared overlap with the decoupled case: $|F(0,\gamma)|^2$. 
We find that $|F(0,\gamma)|^2$ shows relatively large values in the ranges of Eq.~\eqref{eq:QSH_range}, 
indicating that the product of independent QH states continues to give a good approximation to the GS in these ranges. 

The estimated parameter ranges of QSH states in Eqs.~\eqref{eq:QSH_nu1} and \eqref{eq:QSH_range} 
are shown by horizontal bars in Fig.~\ref{fig:phase}. 
These ranges are remarkably wide and even include the case of $g_{\ua\da} \approx g$. 
This sharply contrasts with the case of parallel magnetic fields, 
where the product of two QH states persists only up to 
$g_{\ua\da}/g\approx 0.6$ \cite{FurukawaUeda12} and $0.2$ \cite{Regnault13} for $\nu_\tot=4/3$ and $2$, respectively.  

\subsection{Phase transition at $\gamma=-\pi/4$} \label{sec:collapse}

\begin{figure}
\begin{center}
\includegraphics[width=0.48\textwidth]{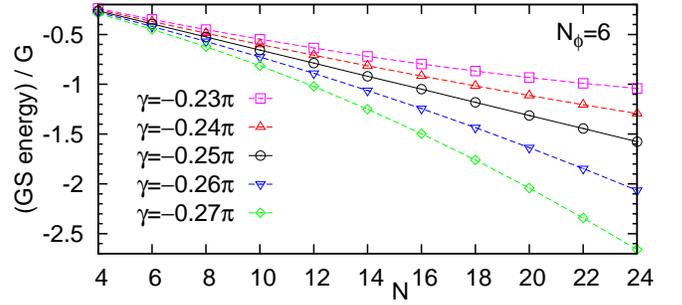}
\end{center}
\caption{(color online)
Ground-state energy as a function of $N$ with $N_\phi=6$ 
for different values of $\gamma$ around the exact transition point $-\pi/4$. 
}
\label{fig:ene_N}
\end{figure}

Here we analyze the phase transition at $\gamma=-\pi/4$, 
which was detected by the peak $\gamma_{c1}(N)$ in the fidelity susceptibility $\chi_F(\gamma)$ in the preceding subsection. 
A unique feature of this transition can be found in the dependence  of the GS energy $E_\mathrm{GS}$ on $N$. 
Figure~\ref{fig:ene_N} displays the numerically calculated $E_\mathrm{GS}(N)$ with $N_\phi=6$ 
at some values of $\gamma$ around $-\pi/4$. 
We find that $E_\mathrm{GS}(N)$ is convex for $\gamma>-\pi/4$ and concave for $\gamma<-\pi/4$. 
We show later in Sec.~\ref{sec:exact_pair_gs} that 
$E_\mathrm{GS}(N)$ is exactly linear at $\gamma=-\pi/4$ as in Eq.~\eqref{eq:EGS_paired} within the LLL approximation. 
This indicates that the compressibility 
\begin{equation}\label{eq:compressibility}
 \kappa= \left[ \frac{N^2}{4\pi R^2} \frac{d^2 E_\mathrm{GS}}{dN^2} \right]^{-1}
\end{equation} 
changes the sign across $\gamma=-\pi/4$ (with a divergence $\kappa\to\pm\infty$ at the transition point). 
The states with $\kappa<0$ in the region $\gamma<-\pi/4$ are thermodynamically unstable and spontaneously contract, 
leading to a collapse of the gas (see the comment in \cite{Comment_collapse}). 
Since the linear behavior of $E_\mathrm{GS}(N)$ at $\gamma=-\pi/4$ [Eq.~\eqref{eq:EGS_paired}] is exact for arbitrary even integer $N>0$, 
$\gamma=-\pi/4$ (i.e., $g_{\ua\da}/g=-1$) gives the exact phase boundary for arbitrary $\nu_\tot$ (within the LLL approximation) as shown in Fig.~\ref{fig:phase}.

\section{Exact results for $g_{\ua\da}=-g$} \label{sec:exact_gs}

We here discuss some exact results at the point $g_{\ua\da}=-g$ within the LLL manifold. 
At this point, the phase transition occurs, and the system collapses for $g_{\ua\da}<-g$ as discussed in Sec.~\ref{sec:collapse}. 
We start by deriving the exact many-body GS with a novel pairing nature in Sec.~\ref{sec:exact_pair_gs}. 
In Sec.~\ref{sec:su11}, we reveal the su(1,1) structure hidden behind this solution, 
and use it to calculate the paring amplitude of the GS. 
In Sec.~\ref{sec:exact_es}, we present exact construction of some excited states, 
and use it to calculate the neutral gap $\Delta_n$. 

\subsection{Exact pairing ground states} \label{sec:exact_pair_gs}


Using the LLL states $\{\psi_{m\alpha}\}$ on a sphere [see Eq.~\eqref{eq:psi_m_ua_da}], the field operator can be expanded  as 
\begin{equation}\label{eq:Psi_expand}
 \Psih_\alpha (\rv) = \sum_{m=-S}^{S} b_{m\alpha} \psi_{m\alpha} (\rv), 
\end{equation}
where the bosonic operators $\{b_{m\alpha}\}$ satisfy the commutation relation 
\begin{equation}
 [b_{m\alpha}, b_{m'\alpha'}^\dagger]=\delta_{mm'}\delta_{\alpha\alpha'}.
\end{equation}
From this expansion, one can derive the equal-position commutation relation 
\begin{equation}
[\Psih_\alpha(\rv),\Psih_\beta^\dagger(\rv)]=\delta_{\alpha\beta} \frac{2S+1}{4\pi S \ell^2}. 
\end{equation}
Using this, the interaction Hamiltonian \eqref{eq:Hint} at the point $g_{\ua\da}=-g$ can be rewritten as
\begin{equation}\label{eq:Hint_solvable}
H_\mathrm{int}
=\frac{g}{2} \int d^2\bm{r} \left[\rho_\uparrow(\bm{r})-\rho_\downarrow(\bm{r})\right]^2
-\frac{g}{\ell^2}\frac{2S+1}{8\pi S}(N_\uparrow+N_\downarrow), 
\end{equation}
where $\rho_\alpha(\rv)=\Psih_\alpha^\dagger(\rv) \Psih_\alpha(\rv)$ is the density operator. 
Thus the GS can be obtained 
by minimizing the squared density difference $\left[\rho_\uparrow(\bm{r})-\rho_\downarrow(\bm{r})\right]^2$ as much as possible 
for every point $\rv$ in the space. 
In fact, one can make this difference vanish everywhere. 
To see it, we introduce the operator 
\begin{equation}\label{eq:Kp}
 K_+ = \int d\rv \Psih_\ua^\dagger (\rv) \Psih_\da^\dagger (\rv), 
\end{equation}
which creates a tightly bound pair of $\ua$ and $\da$ particles uniformly in space. 
Noting $[\Psih_{\ua,\da}(\rv),K_+]=\Psih_{\da,\ua}^\dagger (\rv)$, 
one can show $[\rho_\uparrow(\bm{r})-\rho_\downarrow(\bm{r}),K_+]=0$. 
Therefore, starting from the vacuum $|0\rangle$ and repeatedly operating $K_+$ on it, 
we can make the density difference vanish everywhere. 
The GS for particle numbers $N_\ua=N_\da=N/2$ is thus obtained as
\begin{equation} \label{eq:GS_N}
 |\Psi(N)\rangle \propto K_+^{N/2}|0\rangle , 
\end{equation}
whose energy is given by the last term in Eq.~\eqref{eq:Hint_solvable}:
\begin{equation}\label{eq:EGS_paired}
 E_\mathrm{GS}(N)=-\frac{g}{\ell^2}\frac{2S+1}{8\pi S}N . 
\end{equation}
The linear dependence of the GS energy on $N$ derived here 
leads to a divergence of the compressibility \eqref{eq:compressibility} as discussed in Sec.~\ref{sec:collapse}. 

\subsection{Hidden su(1,1) structure and paring amplitude} \label{sec:su11}

We first reveal the su(1,1) structure hidden in the present model. 
We introduce 
\begin{equation}
\begin{split}
 K_-=K_+^\dagger,~
 K_z=\frac12 (N_\ua+N_\da+2S+1), 
\end{split}
\end{equation}
which together with $K_+$ in Eq.~\eqref{eq:Kp}, satisfy the su(1,1) Lie algebra \cite{Novaes04,Ueda02}:
\begin{equation}
 [K_+,K_-]=-2K_z,~~
 [K_z,K_\pm]=\pm K_\pm.
\end{equation}
The Casimir operator, which is an analogue of the magnitude of the angular momentum in su(2), 
is defined as
\begin{equation}
\begin{split}
 C
 &=K_z^2 - \frac12 (K_+K_- + K_-K_+)\\
 &=K_z(K_z-1) -K_+K_-.
\end{split}
\end{equation}
This operator commutes both with $K_z$ and $K_\pm$. 
A set of basis vectors $\{ |k,n\rangle \}$ for a representation of su(1,1) 
can be chosen to be the simultaneous eigenvectors of $C$ and $K_z$: 
\begin{equation}
\begin{split}
 &C|k,n\rangle = k(k-1) |k,n\rangle,\\
 &K_z |k,n\rangle = (k+n) |k,n\rangle, 
\end{split}
\end{equation}
where the real number $k>0$ is called the Bargmann index and $n$ can be any nonnegative integer. 
In this basis, the operators $K_\pm$ play the roles of raising and lowering $n$ by one:
\begin{subequations}
\begin{align}
 &K_+ |k,n\rangle = \sqrt{(n+1)(2k+n)} ~|k,n+1\rangle,\\
 &K_- |k,n\rangle = \sqrt{n(2k+n-1)} ~|k,n-1\rangle. \label{eq:K-}
\end{align}
\end{subequations}

Since the vacuum $|0\rangle$ is an eigenstate of $C$ and $K_z$ with $k=S+1/2$ and $n=0$, 
the exact GS $|\Psi(N)\rangle$ in Eq.~\eqref{eq:GS_N} can be identified with $|k=S+1/2,n=N/2\rangle$. 
The paring amplitude in $|\Psi(N)\rangle$ is thus calculated by using Eq.~\eqref{eq:K-} as  
\begin{equation}\label{eq:pair_amp}
\begin{split}
 &\frac{1}{(4\pi R^2)^2} \langle \Psi(N)| K_+ K_- |\Psi(N)\rangle \\
 &=\frac{1}{(4\pi \ell^2)^2 S^2} \frac{N}{2} \left( 2S+\frac{N}{2} \right) 
 \longrightarrow \frac{\nu_\tot (2+\nu_\tot) }{(4\pi \ell^2)^2} .
\end{split}
\end{equation}
In the last expression, we take the thermodynamic limit $N\to\infty$ while keeping $\nu_\tot=N/(2S)$ fixed. 
This result indicates that the paring amplitude remains finite in the thermodynamic limit  for any $\nu_\tot>0$ and is a monotonically increasing function of $\nu_\tot$. 


We note that the su(1,1) structure presented above is not limited to a spherical geometry 
but exits in a torus geometry as well; in the latter case, we only have to replace the definition of $K_z$ by $K_z=\frac12 (N_\ua+N_\da+N_\phi)$.  
The paring amplitude can also be calculated in a similar manner, resulting in the same expression as in Eq.~\eqref{eq:pair_amp} in the thermodynamic limit. 

\subsection{Excited states}\label{sec:exact_es}

Here we present exact construction of some excited states. 
To this end, we introduce extended pair creation operators 
\begin{equation}
 P_{JM}^\dagger=\sum_{m_1,m_2} b_{m_1\ua}^\dagger b_{m_2\da}^\dagger \langle S,m_1;S,m_2|J,M\rangle, 
\end{equation}
where $\langle S,m_1;S,m_2|J,M\rangle$ is the Clebsch-Gordan coefficient. 
This operator creates a pair of $\ua$ and $\da$ particles with the total angular momentum $(J,M)$. 
Substituting the expansion \eqref{eq:Psi_expand} into \eqref{eq:Kp}, one can show 
\begin{equation}
 K_+= \sum_{m=-S}^S (-1)^{S-m} b_{m\ua} b_{-m,\da} = \sqrt{2S+1}P_{00}^\dagger. 
\end{equation}
Therefore, the GS $|\Psi(N)\rangle$ in Eq.~\eqref{eq:GS_N} consists of $N/2$ pairs of $\ua$ and $\da$ particles with the angular momentum $(J,M)=(0,0)$. 
Our idea is to break one such pair into a state with higher $J$. 

We start by constructing exact 2-body eigenstates for $N_\ua=N_\da=1$. 
In this case, the eigenstates are given simply by $P_{JM}^\dagger|0\rangle$ 
since the Hilbert space is fully decomposed in terms of $(J,M)$. 
Denoting its eigenenergy by $E_J$ and using the Hamiltonian \eqref{eq:Hint_bbbb} (see Appendix~\ref{app:LLL_sphere}), the eigenequation is written as
\begin{equation}\label{eq:V_eig}
\begin{split}
 &\sum_{m_3,m_4} 2V_{m_1,m_2,m_3,m_4}^{\ua\da} \langle S,m_4,S,m_3|J,M\rangle \\
 &= E_J \langle S,m_1,S,m_2|J,M\rangle 
\end{split}
\end{equation}
where $V_{m_1,m_2,m_3,m_4}^{\ua\da}$ is shown in Eq.~\eqref{eq:V_mmmm_ud}. 
Although we know that this eigenequation should be satisfied because of the spherical symmetry, 
the calculation of $E_J$ is not simple in general 
because the Clebsch-Gordan coefficients become increasingly more complicated for larger $J$. 
Here we perform calculations for $(J,M)=(1,0)$; 
in this case, the Clebsch-Gordan coefficients are simple and their non-zero values are given by 
\begin{equation}
 \langle J,m; J,-m | 1,0\rangle = \sqrt{\frac{3}{S(S+1)(2S+1)}} (-1)^{S-m} m. 
\end{equation}
Using this, the left-hand side of the eigenequation \eqref{eq:V_eig} is calculated as
\begin{equation}\label{eq:V_eig_J1}
\begin{split}
 &\sum_{m_3,m_4} 2V_{m_1,m_2,m_3,m_4}^{\ua\da} \langle S,m_4,S,m_3|1,0\rangle \\
 &=\delta_{m_1+m_2,0} \frac{-g}{4\pi\ell^2} \frac{(2S+1)^2}{S(4S+1)} 
     \sqrt{\frac{3}{S(S+1)(2S+1)}} \\
 &    \times (-1)^{S-m_1} \sum_{m_4} \frac{C(2S,S+m_1) C(2S,S+m_4)}{C(4S,2S+m_1+m_4)} m_4\\
 &= -\frac{g}{4\pi\ell^2} \frac{2S+1}{S+1} \langle S,m_1,S,m_2|1,0\rangle ,
\end{split}
\end{equation}
where we have used the following identity for the binomial coefficient $C(\cdot,\cdot)$:
\begin{equation}
\begin{split}
 &\sum_{m_4=-S}^S \frac{C(2S,S+m_1) C(2S,S+m_4)}{C(4S,2S+m_1+m_4)} m_4 \\
 &= \frac{S(4S+1)}{(S+1)(2S+1)}m_1. 
\end{split}
\end{equation}
The two-body eigenenergy $E_1(2)$ can then be read out from Eq.~\eqref{eq:V_eig_J1} as
\begin{equation}
 E_1(2)=-\frac{g}{4\pi\ell^2} \frac{2S+1}{S+1} .
\end{equation}

Excited states for particle numbers $N_\ua=N_\da=N/2$ can be obtained 
by repeated operations of $K_+$ on the two-body excited state: 
\begin{equation}\label{eq:Psi_N_JM}
|\Psi_{JM} (N)\rangle \propto K_+^{\frac{N}{2}-1} P_{JM}^\dagger |0\rangle.
\end{equation} 
Using 
\begin{equation}
 [H,K_+] = - \frac{g}{4\pi\ell^2} \frac{2S+1}{S} K_+, 
\end{equation}
the eigenenergy for Eq.~\eqref{eq:Psi_N_JM} is calculated as
\begin{equation}
 E_J(N) = E_J (2) - \frac{g}{4\pi\ell^2} \frac{2S+1}{S} \left( \frac{N}{2}-1 \right). 
\end{equation}
Therefore, the excitation gap to the $J=1$ state is obtained as
\begin{equation}\label{eq:gap_J_1}
 E_1(N)-E_\mathrm{GS}(N)=E_1(2)+\frac{g}{4\pi\ell^2} \frac{2S+1}{S} 
 =\frac{g}{4\pi\ell^2} \frac{2S+1}{S(S+1)}. 
\end{equation}
Although we do not have a rigorous argument that $|\Psi_{1M}(N)\rangle$ defined by Eq.~\eqref{eq:Psi_N_JM} should be the first excited state, 
we have confirmed that the numerical data of the neutral energy gap [the area of the circles in Fig.~\ref{fig:gsLtri}(b)] 
agrees with this expression. 
This gap vanishes for $S\to\infty$, indicating that the system is gapless in the thermodynamic limit. 

\section{Summary and outlook}\label{sec:conclude}

In this paper, we have determined the global phase diagram of two-component Bose gases in antiparallel magnetic fields as shown in Fig.~\ref{fig:phase}. 
We have found that QSH states composed of two nearly independent QH states 
are remarkably stable and persist even for $g_{\ua\da}$ up to as large as $g$. 
This sharply contrasts with the case of parallel magnetic fields, 
where SU(2)-symmetric QH states (with high entanglement between the two components) 
emerge for $g_{\ua\da}\approx g$ at $\nu_\tot=4/3$ \cite{Hormozi12,Grass12,FurukawaUeda12,Wu13} and $2$ \cite{Senthil13,FurukawaUeda13,Wu13,Regnault13}. 
In spite of this marked difference between the cases of parallel and antiparallel magnetic fields for small $\nu_\tot$, 
we have shown within the GP mean-field theory that for large $\nu_\tot(\gg 1)$, 
the two cases show the same vortex phase diagrams. 
We have also shown (within the LLL approximation) that the vertical line $g_{\ua\da}/g=-1$ in Fig.~\ref{fig:phase} gives the exact critical line  beyond which the system collapses. 
Along this line, we have obtained the exact many-body GS with a novel paring nature. 

The present study of a simple model could serve as a useful reference point 
for studying the interplay between intracomponent and intercomponent interactions in more general spin Hall systems. 
The possibilities of novel time-reversal invariant topological states (beyond simple product states)  
for strong intercomponent interactions in 2D fermionic and bosonic spin Hall systems 
have been proposed in Refs.~\cite{Bernevig06,Levin12,Santos11,Chen12,Lu12}.  
Our results however have demonstrated that in the present system, simple products of QH states are remarkably stable, 
and increasing the intercomponent coupling does not produce any non-product-type topological state. 
It would be interesting to investigate whether other perturbations such as the introduction of an optical lattice or longer-range interactions 
can produce a non-product-type topological state. 
Strongly correlated physics in spin-orbit-coupled systems has also received growing interest
in the studies of heavy transition-metal compounds such as iridates \cite{Witczak14}. 
Parallel studies of cold-atom and solid-state systems would be beneficial 
for a unified understanding of correlated quantum phenomena in spin Hall systems. 

\bigskip 

The authors thank M. A. Cazalilla and Y. Horinouchi for useful discussions. 
This work was supported by 
KAKENHI Grant Nos.~25800225, 22340114, and 26287088 from the Japan Society for the Promotion of Science, 
and by a Grant-in-Aid for Scientific Research on Innovation Areas "Topological Quantum Phenomena" (KAKENHI Grant No.~22103005) 
and the Photon Frontier Network Program from MEXT of Japan.

{\it Note added.} 
After completion of this work, we became aware of an independent work by Repellin {\it et al.} \cite{Repellin14}, 
where the stability of two coupled bosonic Laughlin states is investigated in lattice models. 
Although their models are different from ours, 
they have also reached the conclusion that the product of two Laughlin states with opposite chiralities 
persists over a wider range of the intercomponent coupling than the case of the same chiralities. 

\appendix

\section{Hamiltonian in the lowest-Landau-level basis on a sphere}\label{app:LLL_sphere}

Here we describe some basic facts about the lowest-Landau-level (LLL) basis on a sphere \cite{Haldane83,Fano86} 
in our time-reversal invariant setting, 
and derive the representation of the interaction Hamiltonian in this basis. 
We also briefly explain the numerical methods that we used for diagonalizing this Hamiltonian. 

We consider the setting described in Sec.~\ref{sec:sphere}. 
The single-particle Hamiltonian ${\cal K}_\alpha$ for the spin state $\alpha(=\ua,\da)$ on the sphere is given by
\begin{equation}
 {\cal K}_\alpha = \frac{1}{2M} \left[ \ev_r \times (\pv-q\Av_\alpha) \right]^2=\frac{\Lambdav_\alpha^2}{2MR^2}, 
\end{equation}
with
\begin{align}
 \Av_{\ua,\da} &= \mp N_\phi \frac{2\pi\hbar}{q} \frac{\cot\theta}{4\pi r} \ev_\phi, \\
 \Lambdav_\alpha &=\rv \times (\pv-q\Av_\alpha).
\end{align}
One can show that $\Lv_{\ua,\da}=\Lambdav_{\ua,\da} \mp \hbar S\ev_r$ obey the standard algebra of an angular momentum; 
henceforth we simply call $\Lv_{\ua,\da}$ the angular momentum. 
It follows from the relation $\Lambdav_\alpha^2=\Lv_\alpha^2-\hbar^2 S^2$ 
that the eigenvalues of $\Lambdav_\alpha^2$ are given by $\hbar^2 [l(l+1)-S^2]$, 
where $l$ is the magnitude of the angular momentum $\Lv_\alpha$. 
Because $\Lambdav_\alpha^2\ge 0$, the minimum of $l$ is given by $l=S$. 

The LLL on a sphere is thus given by the states with the angular momentum $l=S$. 
Their wave functions are given by
\begin{subequations}\label{eq:psi_m_ua_da}
\begin{align}
 &\psi_{m\ua} (\rv) =\frac1R \left[ \frac{2S+1}{4\pi} C(2S,S-m) \right]^{\frac12} (v^*)^{S+m} (-u^*)^{S-m}, \\
 &\psi_{m\da} (\rv) = \frac1R \left[ \frac{2S+1}{4\pi} C(2S,S+m) \right]^{\frac12} u^{S+m} v^{S-m}, 
\end{align}
\end{subequations}
where $m=-S,-S+1,\dots,S$ is the $z$-component of the angular momentum, 
$C(\cdot,\cdot)$ is the binomial coefficient, and
\begin{equation}
 u=\cos \left( \frac{\theta}{2} \right) e^{i\phi/2},~~
 v=\sin \left( \frac{\theta}{2} \right) e^{-i\phi/2}. 
\end{equation}
These wave functions satisfy
\begin{align}
 \psi_{m\ua} = (-1)^{S-m} \psi_{-m,\da}^*,~~\psi_{m\da}= (-1)^{S+m} \psi_{-m,\ua}^*. 
\end{align}

Substituting the mode expansion \eqref{eq:Psi_expand} into the interaction Hamiltonian \eqref{eq:Hint}, we obtain
\begin{equation}\label{eq:Hint_bbbb}
 H_\mathrm{int}
 = \sum_{\alpha,\beta} \sum_{m_1,m_2,m_3,m_4} V_{m_1 m_2 m_3 m_4}^{\alpha\beta} b_{m_1\alpha}^\dagger b_{m_2\beta}^\dagger b_{m_3\beta} b_{m_4\alpha},
\end{equation}
where
\begin{equation}\label{eq:V_mmmm}
\begin{split}
 &V_{m_1 m_2 m_3 m_4}^{\alpha\beta} \\
 &= \frac{g_{\alpha\beta}}{2} \int d^2 \rv\psi_{m_1\alpha}^*(\rv) \psi_{m_2\beta}^*(\rv) \psi_{m_3\beta}(\rv) \psi_{m_4\alpha}(\rv). 
\end{split}
\end{equation}
Using Eq.~\eqref{eq:psi_m_ua_da}, the coefficients \eqref{eq:V_mmmm} are calculated as
\begin{align}
V_{m_1 m_2 m_3 m_4}^{\alpha\alpha} 
&=\delta_{m_1+m_2,m_3+m_4} \frac{g}{8\pi \ell^2} \notag\\
&\times \frac{(2S+1)^2}{S(4S+1)} \frac{\left( \prod_i C(2S,S+m_i) \right)^{1/2}}{ C(4S,2S+m_1+m_2) },\\
V_{m_1 m_2 m_3 m_4}^{\ua\da} 
&=\delta_{m_1+m_2,m_3+m_4} (-1)^{2S-m_1-m_4} \frac{g_{\ua\da}}{8\pi \ell^2} \notag\\
&\times \frac{(2S+1)^2}{S(4S+1)} \frac{\left( \prod_i C(2S,S+m_i) \right)^{1/2}}{ C(4S,2S+m_1-m_3) }. \label{eq:V_mmmm_ud}
\end{align}

The total angular momentum $\Lv^\tot$ can be expressed in the second-quantized form as 
\begin{equation}
\begin{split}
 L_+^\tot &= L_x^\tot+i L_y^\tot \\
 &= \sum_{m,\alpha} \sqrt{(S-m)(S+m+1)} b_{m+1\alpha}^\dagger b_{m\alpha},\\
 L_-^\tot &= L_x^\tot-i L_y^\tot =(L_+^\tot)^\dagger,\\
 L_z^\tot &= \sum_{m,\alpha} m b_{m\alpha}^\dagger b_{m\alpha}.
\end{split}
\end{equation}
These operator commute with the Hamiltonian \eqref{eq:Hint_bbbb}. 
Therefore, the eigenstates of Eq.~\eqref{eq:Hint_bbbb} can be classified 
by the magnitude $L$ and the $z$-component eigenvalue $L_z$ of $\Lv^\tot$. 

We performed exact diagonalization calculations of the Hamiltonian \eqref{eq:Hint_bbbb}. 
Restriction to the subspace of the fixed particle numbers $(N_1,N_2)$ and the fixed projected total angular momentum $L_z$ 
can easily be implemented by an appropriate choice of the Fock basis $\{ |\{n_{m\alpha}\}\rangle \}$, 
where $n_{m\alpha}$ is the eigenvalue of $b_{m\alpha}^\dagger b_{m\alpha}$. 
Restriction to specific $L=L_\mathrm{target}$ can be done by first choosing the sector with $L_z=L_\mathrm{target}$ 
and then adding to the Hamiltonian the term $\lambda L_-^\tot L_+^\tot$, where $\lambda$ is a positive constant. 
This term shifts the entire eigenspectrum by $\lambda [L(L+1)-L_z(L_z+1)]$. 
The states with $L>L_\mathrm{target}$ can therefore be eliminated from the low-energy subspace of our interest 
by taking sufficiently large $\lambda>0$. 
We used a LAPACK full diagonalization routine, 
an ARPACK routine \cite{ARPACK} of the implicitly restarted Lanczos method, 
and the standard Lanczos method (storing only a minimum of two Lanczos vectors at each iteration)
for small, medium, and large system sizes. 
The second method requires a larger memory space for Lanczos vectors than the third, 
but is more stable against the presence of eigenenergy degeneracies. 
In fact, the present time-reversal invariant Hamiltonian often involves such degeneracies, 
and the standard Lanczos method sometimes (but not always) does not converge properly 
(for all presented data calculated by this method, we have checked sufficient convergence). 
The addition of the $\lambda$ term for the $L$-restriction mentioned above 
makes the convergence even worse (because the matrix to be diagonalized becomes less sparse); 
we therefore imposed this restriction only with LAPACK and ARPACK routines. 
Even when we did not add the $\lambda$ term, 
we could determine the $L$ value of each eigenstate by operating $(\Lv^\tot)^2$ on it. 
In Fig.~\ref{fig:gsLtri}, for example, we used the standard Lanczos method without the $L$-restriction for large system sizes, 
but could still determine the $L$ value of the ground state in this way. 






\begin{thebibliography}{99}
\bibitem{Comment_Fialko14}
Contrary to our results, Ref.~\cite{Fialko14} found gapless-like features for $g_{\uparrow\downarrow}=g$. 
However, the results of Ref.~\cite{Fialko14} are based on small system sizes ($N=4$ and $6$) 
and non-uniform geometries, and more likely to suffer from finite-size effects.
\bibitem{Comment_collapse}
We note that within the lowest-Landau-level (LLL) approximation, 
a genuine collapse with a divergence in the density occurs only in the thermodynamic limit. 
In Sec.~\ref{sec:collapse}, we discuss within this approximation 
that for $g_{\ua\da}<-g$, the compressibility \eqref{eq:compressibility} is negative, 
and thus the gas spontaneously contracts. 
However, the gas cannot shrink into a point 
because of the non-commutativity of the coordinates within the LLL manifold 
($[x,y]=-i\ell^2/2$, where $\ell$ is the magnetic length). 
Therefore, for finite $N$, the maximal achievable density is of the order of $N/\ell^2$.
\end{thebibliography}
\end{document}